\documentclass[a4paper]{article}

\usepackage{INTERSPEECH2021}

\title{Audiovisual transfer learning for audio tagging and sound event detection}
\name{Wim Boes$^1$, Hugo Van hamme$^1$}
\address{$^1$ESAT, KU Leuven}
\email{wim.boes@esat.kuleuven.be, hugo.vanhamme@esat.kuleuven.be}

\usepackage{xcolor}
\usepackage{multirow}
\usepackage{pifont}
\usepackage{tikz}
\usetikzlibrary{shapes.misc}
\tikzset{
 draw/.append style={font=\normalsize},
 font={\fontsize{8pt}{12} \fontfamily{libertine} \selectfont},
 cross/.style={
 cross out, draw, minimum size=2*(#1-\pgflinewidth), inner sep=0pt, outer sep=0pt
 }
}

\begin{document}

\maketitle
 
\begin{abstract}
We study the merit of transfer learning for two sound recognition problems, i.e., audio tagging and sound event detection. Employing feature fusion, we adapt a baseline system utilizing only spectral acoustic inputs to also make use of pretrained auditory and visual features, extracted from networks built for different tasks and trained with external data.

We perform experiments with these modified models on an audiovisual multi-label data set, of which the training partition contains a large number of unlabeled samples and a smaller amount of clips with weak annotations, indicating the clip-level presence of 10 sound categories without specifying the temporal boundaries of the active auditory events.

For clip-based audio tagging, this transfer learning method grants marked improvements. Addition of the visual modality on top of audio also proves to be advantageous in this context.

When it comes to generating transcriptions of audio recordings, the benefit of pretrained features depends on the requested temporal resolution: for coarse-grained sound event detection, their utility remains notable. But when more fine-grained predictions are required, performance gains are strongly reduced due to a mismatch between the problem at hand and the goals of the models from which the pretrained vectors were obtained. 
\end{abstract}

\section{Introduction}
\label{sect:intro}

Sound recognition tasks have recently gained a lot of interest in the machine learning community. Reasons for the popularity surge of these topics include but are not limited to the release of large-scale data sets such as Audio Set~\cite{AudioSet} and the organization of multiple editions of the Detection and Classification of Acoustic Scenes and Events (DCASE) challenge~\cite{DCASE2016, DCASE2017proceedings, DCASE2018proceedings, DCASE2019proceedings, DCASE2020proceedings}.

This work deals with two important and related classification subjects in this field of machine learning. The first is commonly referred to as audio tagging (AT): for this task, the goal is to determine which types of sounds appear in given sound clips. 

The objective of the second task, which is usually called sound event detection (SED), is to generate transcriptions of audio recordings: instead of only needing to provide tags indicating whether or not particular sounds appear in clips as is the case for audio tagging models, sound event detection systems are also required to predict the temporal boundaries of all auditory events to some predefined degree of accuracy. 

Because of this extra requirement, it is unquestionably more difficult to make capable sound event detection systems than it is to construct effective audio tagging models, especially when it comes to the data used for training: it is easier to label audio recordings only in terms of the presence of sound categories than it is to also include annotations of the temporal boundaries of events. Depending on the needs of the targeted application, it could thus be wiser to build audio tagging systems due to the availability of more data and/or data with better labeling.

The fourth task of DCASE 2017~\cite{DCASE2017proceedings} was dedicated to audio tagging. The winning system~\cite{gatedconv} was a convolutional recurrent neural network (CRNN) with learnable gated linear units.

One subtask of each of the DCASE 2018, DCASE 2019 and DCASE 2020 challenges~\cite{DCASE2018proceedings, DCASE2019proceedings, DCASE2020proceedings} dealt with sound event detection. Many of the best-performing systems employed some neural architecture such as a CRNN~\cite{gatedconv, dcase2018winner, dcase2019winner}, a transformer~\cite{transformers, dcase2020winner} or a conformer~\cite{dcase2020winner, conformers} in conjunction with the mean teacher training principle~\cite{meanteacher} and in some cases also data augmentation techniques such as SpecAugment~\cite{specaugment} and mixup~\cite{mixup}. 

A majority of the data available for the training of systems for the aforementioned challenge subtasks consisted of audio samples originated from YouTube. In this context, it is certainly feasible to also extract visual information and subsequently incorporate this extra modality into sound recognition models. Previous works~\cite{weakly, audiovisual} have demonstrated that doing so can provide considerable improvements for the audio tagging task. However, to the best of our knowledge, this idea has not yet been explored for the problem of sound event detection.

The foregoing challenge subtasks forbade the use of data that was not explicitly provided to the participants. As a result, the systems developed for these problems typically employed some type of spectral features (e.g., log mel magnitude spectrograms) directly extracted from the given sound recordings. In a more general setting, it is also possible to utilize features computed by externally pretrained neural networks to incorporate the knowledge they capture. This type of transfer learning has been used in previous works: pretrained auditory and visual features have been employed to perform audio tagging~\cite{weakly, audiovisual}. However, these research efforts do not include a complete comparison of models with and without pretrained features and to the best of our knowledge, no examination of this phenomenon has been performed for sound event detection.

The main contribution of this work is to fill in the gaps discussed in the two preceding paragraphs: for both audio tagging and sound event detection, we make a direct comparison between models utilizing (non-pretrained) spectral features and pretrained auditory features. Furthermore, we investigate the advantage of using multimodal, audiovisual information for both of these sound recognition tasks.

To this end, the following approach is taken: we start from a high-performing baseline utilizing only spectral auditory features that can be employed for both audio tagging and sound event detection. Making use of a straightforward feature fusion method, this system is extended to also be able to handle pretrained auditory and visual features. We then analyze the sound recognition results obtained by these adapted models in a number of different experimental settings to infer conclusions.

In Section~\ref{sect:baseline}, we describe the baseline system. Next, in Section~\ref{sect:fusion}, the employed feature fusion method is explained. In Section~\ref{sect:setup}, we delve into the experimental setup. Then, in Section~\ref{sect:results}, we analyze the results of the performed experiments and finally, we draw a conclusion in Section~\ref{sect:conclusion}.

\begin{figure*}[!ht]
\centering
\includegraphics[scale=1]{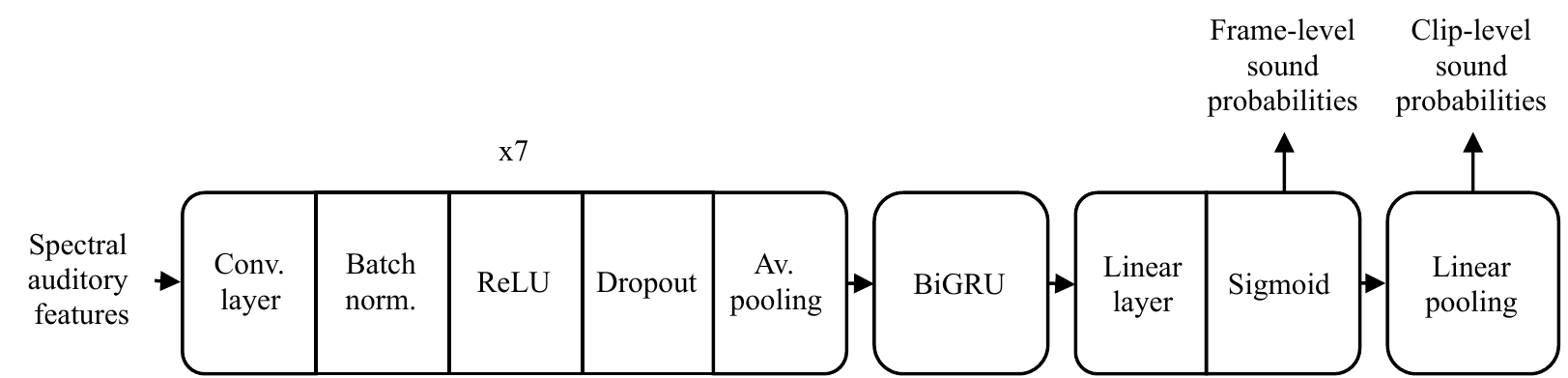}
\caption{Architecture of the baseline system.}
\label{fig:baseline}
\end{figure*}

\section{Baseline}
\label{sect:baseline}

In this section, we describe the baseline for both of the considered sound recognition tasks elaborated on in Section~\ref{sect:intro}, namely audio tagging and sound event detection. The system is a slightly adapted version of the winning CRNN model~\cite{dcase2018winner} of task 4 of the DCASE 2018 challenge~\cite{DCASE2018proceedings}. 

\subsection{Architecture}

The baseline architecture is shown in Figure~\ref{fig:baseline}.

The input to the network is a spectral (time-frequency) map of the considered audio recording. The number of frequency bins used is fixed and equal to 128.

The first component of the architecture is a convolutional neural network (CNN), consisting of a sevenfold repetition of a stack of five layers: first, a convolution is executed, followed by batch normalization~\cite{batchnorm} and use of the non-linear rectified linear unit (ReLU). Next, during training, dropout~\cite{dropout} with a rate of 33\% is applied and finally, average pooling is performed. Relevant hyperparameters of the convolutional and pooling layers are given in Table~\ref{tab:cnnhyper} and Table~\ref{tab:poolhyper}. For the listed kernel sizes and strides, the first and second numbers of each tuple relate to the time and frequency axes respectively. At the end of the last stack, the frequency-related dimension of the original spectral map has been reduced to one and is therefore discarded. 

\begin{table}[!ht]
\caption{Hyperparameters of convolutional layers in CNN}
\label{tab:cnnhyper}
\centering
\begin{tabular}{@{}lccc@{}}
\toprule
\textbf{Stack} & \textbf{Kernel size} & \textbf{Stride} & \textbf{Output channels} \\
\midrule
0 & (3, 3) & (1, 1) & 16 \\
1 & (3, 3) & (1, 1) & 32 \\
2 & (3, 3) & (1, 1) & 64 \\
3-4-5-6 & (3, 3) & (1, 1) & 128 \\
\bottomrule
\end{tabular}
\end{table}

\begin{table}[!ht]
\caption{Hyperparameters of average pooling layers in CNN}
\label{tab:poolhyper}
\centering
\begin{tabular}{@{}lc@{}}
\toprule
\textbf{Stack} & \textbf{Kernel size = stride} \\
\midrule
0-1 & (2, 2) \\
2-3-4-5-6 & (1, 2) \\
\bottomrule
\end{tabular}
\end{table}

After the CNN, a two-layered bidirectional gated recurrent unit (BiGRU) with a hidden size of 128 is employed to model temporal dependencies. The output of this module is processed by a linear layer followed by application of the sigmoid non-linearity to obtain multi-label frame-level probabilities, indicating per temporal frame which sound categories are (in)active.

Finally, the frame-level sound probabilities are aggregated to get clip-level audio event probabilities by performing linear pooling, which can be mathematically expressed as follows:

\begin{equation}
p_c\left(n\right) = \sum\limits_{t=0}^{T-1} p_f\left(n, t\right) \cdot \left(\dfrac{p_f\left(n, t\right)}{\sum\limits_{t'=0}^{T-1} p_f\left(n, t'\right)}\right)
\label{eq:linearpool}
\end{equation}

In Equation~\eqref{eq:linearpool}, $p_c\left(n\right)$ denotes the clip-level probability for sound class $n$ and $p_f\left(n, t\right)$ is the frame-level probability for the relevant category at time $t$ (of a maximum of $T$ temporal frames). We employ this pooling function due to its simplicity and the theoretical advantages it provides over other aggregation methods as explained in previous research efforts~\cite{poolcomparison}.

\subsection{Postprocessing of probabilities}

To convert the clip-level and frame-level probabilities into binary decisions, we employ a simple, fixed threshold of 0.5.

After applying the threshold, the frame-level decisions are additionally passed through a median filtering operation with a window of 7 temporal frames, corresponding to about 0.5 seconds, to smoothen the predictions. 

\subsection{Mean teacher training method}

As elaborated upon in Section~\ref{sect:data}, the training data utilized in this research project is heterogeneously annotated: some of the samples are weakly annotated in the sense that clip-level information is available, the rest is unlabeled. To deal with this phenomenon, the mean teacher principle~\cite{meanteacher} can be employed.

In the mean teacher framework, two models typically called the student and the teacher are utilized. Architecturally, they are identical, but their weights are updated differently.

Training of the student happens regularly: a differentiable loss function is minimized by an algorithm such as Adam~\cite{adam}. However, the teacher is updated differently: its parameters are computed as the exponential moving average of the student model weights with a multiplicative decay factor of 0.999 per training iteration, hence the name of the method.

The loss used to train the student consists of multiple terms. The first is a clip-level binary cross entropy function, which can only be calculated for weakly annotated data samples. The two others are mean-squared error consistency costs between respectively the clip-level and frame-level output probabilities of the student and teacher models, which are also computed for the unlabeled examples. In this project, the classification and consistency components are summed with weights one and two respectively to obtain the final optimization objective. 

\section{Feature fusion}
\label{sect:fusion}

In this section, we explain how the baseline described in Section~\ref{sect:baseline} is extended to deal with pretrained auditory and visual features on top of (non-pretrained) spectral auditory features.

\begin{figure}[!ht]
\centering
\includegraphics[scale=1]{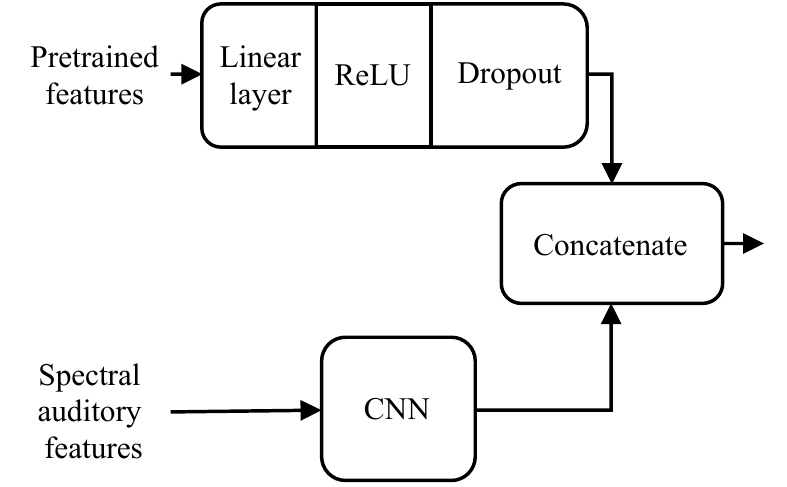}
\caption{Example of employed feature fusion method.}
\label{fig:fusion}
\end{figure}

The used pretrained features are first passed through a linear layer to project them to embeddings with a predefined size. The ReLU non-linearity is used afterwards and during training, dropout~\cite{dropout} with a probability of 33\% is also applied. 

The outputs of these operations can be concatenated among themselves and/or with the 128-dimensional convolutional embeddings extracted from the spectral auditory features before passing the combination on to a bidirectional gated recurrent unit as explained in more detail in Section~\ref{sect:baseline}.

The optimal dimensionalities for the pretrained embeddings were determined experimentally. When used together with non-pretrained spectral acoustic features, the best-performing sizes turn out to be 16 and 4 for the pretrained auditory and visual embeddings respectively. Otherwise, when utilized in a stand-alone manner, the optimal sizes are equal to 64 and 16. 

\section{Experimental setup}
\label{sect:setup}

\subsection{Data}
\label{sect:data}

We work with a modified version of the multi-label data set for problem 4 of DCASE 2018~\cite{DCASE2018proceedings}, which mainly consists of audio files with a length of 10 seconds but also some shorter ones. Unlike the recordings used in later iterations of this task, all clips originate from YouTube, and thus manual downloads allow us to also obtain corresponding visual streams. Regrettably, some videos could not be acquired due to availability issues: about 10\% of examples of the original set had to be discarded.

The training partition holds 1439 samples with weak, clip-level labels of which of 10 possible sounds occur and 13017 unannotated (in-domain) clips. The validation and test sets consist of 259 and 779 samples respectively and are strongly annotated: time boundaries of all auditory events are included.

\subsection{Feature extraction and preprocessing}

To compute spectral auditory features, we first resample the audio clips to 22.1 kHz and perform peak amplitude normalization. Then, logarithmic mel magnitude spectrograms with 128 frequency bins are extracted using a window size of 2048 samples and a hop length of 366 samples. For a recording of 10 seconds, this results in 604 temporal frames. Finally, per-frequency bin standardization is performed. 

To obtain pretrained auditory features, we first resample the sound recordings to 16 kHz and extract log mel magnitude spectrograms with 128 frequency bins using a window size of 400 samples and a hop length of 160 samples. Next, they are divided into chunks of 960 ms with an overlap of 900 ms. These segments are then fed to vggish~\cite{vggish}, a convolutional model for video tag classification based on audio, pretrained using a preliminary version of the YouTube-8M data set~\cite{YouTube8M}. The 128-dimensional outputs of the last feedforward layer of this network are used as pretrained auditory features.

To extract pretrained visual features, we first sample images from the visual streams at a rate of about 15 fps. The resulting frames are then passed through VGG16~\cite{VGG16}, a convolutional network for image classification, pretrained on the ImageNet data set~\cite{ImageNet}. The 4096-dimensional outputs of the last feedforward layer of this model are used as pretrained visual features.

The spectral auditory features and the pretrained features are compatible as it pertains to the feature fusion method outlined in Section~\ref{sect:fusion}: for a 10-second clip, the number of temporal frames is equal to 151 for both the spectral and the pretrained embeddings, hence concatenation is safely possible.

\subsection{Training and evaluation}

All of the models used in the experiments described in this work are trained and evaluated using the PyTorch toolkit~\cite{pytorch}.

\subsubsection{Training hyperparameters}

All systems are trained for 200 epochs. Per epoch, 250 batches of 48 samples are fed to the networks. Each batch contains 24 examples originating from the weakly annotated training set, while the remaining 24 are unlabeled.

We employ Adam~\cite{adam} to update the weights of the student models as explained in Section~\ref{sect:baseline}. Learning rates are ramped up in an exponential manner from 0 to 0.001 for the first 12500 optimization steps. Afterwards, they decay multiplicatively at a rate of 0.99995 per training iteration.

\subsubsection{Evaluation metrics}

For evaluation, we employ some measures commonly used in the relevant subtasks of the DCASE challenges.

For audio tagging, we quantify the performance of the systems by calculating micro-averaged (clip-based) F1 scores~\cite{metrics}. 

For sound event detection, we use two different measures, the first one being the segment-based (micro-averaged) F1 score based on chunks of 1 s~\cite{metrics}. The second metric is the macro-averaged event-based F1 metric with a tolerance of 200 ms for onsets and a collar of 20\% of the lengths of the audio events (up to a maximum of 200 ms) for offsets~\cite{metrics}.

All scores are computed using the output probabilities of the student models obtained after the final training epoch.

\section{Experimental results}
\label{sect:results}

In this section, we analyze the scores obtained by the considered CRNN models using different types of input features, as elaborated upon in Section~\ref{sect:baseline} and Section~\ref{sect:fusion}. We report the metrics mentioned in Section~\ref{sect:setup} after applying the following procedure to ensure the stability and reliability of the results: for every experimental setting, we train 20 systems with independent initializations, select the 5 models with the best performance on the validation partition of the employed data collection and finally, we average the scores they achieve on the test set. 

\begin{table*}[!ht]
\caption{Audio tagging and sound event detection scores of CRNN models}
\label{tab:perf}
\centering
\begin{tabular}{@{}lllccc@{}}
\toprule
\multirow{2}{*}{\begin{tabular}{@{}l@{}} \textbf{Spectral auditory} \\ \textbf{features} \end{tabular}} &
\multirow{2}{*}{\begin{tabular}{@{}l@{}} \textbf{Pretrained auditory} \\ \textbf{features} \end{tabular}} &
\multirow{2}{*}{\begin{tabular}{@{}l@{}} \textbf{Pretrained visual} \\ \textbf{features} \end{tabular}} & 
\multirow{2}{*}{\begin{tabular}{@{}c@{}} \textbf{Clip-based} \\ \textbf{micro-av. F1 score} \end{tabular}} &
\multirow{2}{*}{\begin{tabular}{@{}c@{}} \textbf{Segment-based} \\ \textbf{micro-av. F1 score} \end{tabular}} & 
\multirow{2}{*}{\begin{tabular}{@{}c@{}} \textbf{Event-based} \\ \textbf{macro-av. F1 score} \end{tabular}}\\\\
\midrule
\checkmark & --- & --- & 76.22\% & 70.09\% & 33.03\% \\
\checkmark & --- & \checkmark & 80.72\% & 73.81\% & 32.58\% \\
\checkmark & \checkmark & --- & 81.03\% & 74.74\% & 33.40\% \\
\checkmark & \checkmark & \checkmark & 83.72\% & 76.86\% & 32.65\% \\
\midrule
--- & --- & \checkmark & 61.60\% & 51.83\% & 15.68\%\\
--- & \checkmark & --- & 78.17\% & 70.20\% & 18.83\%\\
--- & \checkmark & \checkmark & 80.04\% & 71.01\% & 22.31\%\\
\bottomrule
\end{tabular}
\end{table*}

Table~\ref{tab:perf} contains the F1 scores achieved by the CRNN models using different combinations of input features.

\subsection{Audio tagging}

As explained in Section~\ref{sect:setup}, the performance of the models with regard to audio tagging is quantified by clip-based F1 scores.

For this task, the employments of pretrained auditory and visual features on top of spectral inputs grant absolute improvements of 4.81\% and 4.50\% respectively. Utilizing both types of pretrained features, this gain is further boosted to 7.50\%. In this situation, the benefit of transfer learning is clear: carrying over knowledge learned by models built for different purposes and trained with a lot of external data provides a serious advantage. 

The models using only pretrained auditory features interestingly achieve better performance than systems exclusively utilizing spectral features. The information inherent to these two different types of inputs also appears to be complementary, as combining them leads to a noticeable improvement. 

Visual inputs are also shown to be useful for audio tagging: when used on their own, they accomplish relatively subpar results, but the multimodal, audiovisual models using pretrained visual features on top of auditory features decisively outperform the unimodal acoustic systems.

\subsection{Sound event detection}

As outlined in Section~\ref{sect:setup}, the performance of the models with regard to the task of sound event detection is evaluated by two different metrics: the micro-averaged segment-based and macro-averaged event-based F1 scores. The former quantifies the systems in terms of their ability to generate coarse-grained transcriptions of auditory events, while the latter assesses their performance concerning fine-grained temporal predictions.

The results for coarse-grained sound event detection appear to be wholly in aligment with the conclusions for the audio tagging task formulated before: employing pretrained auditory and visual features in conjunction with spectral inputs is beneficial in terms of the achieved segment-based F1 scores.

However, for more fine-grained sound event detection, the event-based F1 scores achieved by the models with different types of inputs show entirely different trends.

Utilizing pretrained features on top of spectral inputs does not yield (notable) gains in this context. A plausible reason for this phenomenon is the mismatch between the task at hand and the objectives of the systems from which these vectors were extracted: as outlined in Section~\ref{sect:setup}, the pretrained auditory and visual models were designed for problems without time-related aspects. So, when it comes to making intricate predictions about temporal boundaries of auditory events, their power is limited.

There is an additional reason for the performance degradation caused by the pretrained visual features: solely audio was taken into account during the selection and annotation of samples of the employed data collection. There are no guarantees about the quality of the corresponding visual component or the level of synchronization between the two considered modalities, which clearly translates into a detrimental effect.

When it comes to the models using only pretrained features, we find that they severely underachieve for fine-grained sound event detection. The event-based F1 scores do not drop to zero however: upon manual inspection, we find that these systems still perform well for audiovisual samples that do not really require temporal differentiation, such as an example containing auditory events which are active for the full length of the clip. 

\section{Conclusions}
\label{sect:conclusion}

We investigated transfer learning for audio tagging and sound event detection. More specifically, we employed feature fusion to extend a CRNN model utilizing spectral inputs to also use pretrained auditory and visual vectors, computed by systems designed for other problems and trained with different data.

To train and evaluate these systems, we used an audiovisual multi-label data set. The training samples thereof were partially unannotated and partially weakly labeled, in the sense that clip-level activities of 10 sound categories were indicated.

For audio tagging, we showed that the feature fusion procedure provided significant performance gains. In addition, we demonstrated that spectral and pretrained acoustic features are equally valuable and complementary. Lastly, we illustrated the merit of including another modality (i.e., vision) besides audio.

For sound event detection, results depended on the desired level of granularity: for predicting sounds and making fairly rough estimates of their time boundaries, pretrained auditory and visual features still bestowed substantial improvements. On the other hand, performance boosts for more precise temporal predictions were less convincing. The most likely explanation for this outcome is the lack of time-related aspects in the tasks for which the relevant pretrained models were optimized.

For future research related to fine-grained sound event detection, it could thus be worthwhile to look into pretrained features computed by models constructed for problems with temporal facets. In general, it would also be interesting to analyze more complex transfer learning techniques, possibly using other architectures and data with different properties.

\section{Acknowledgements}

This work is supported by a PhD Fellowship of Research Foundation Flanders (FWO-Vlaanderen).

\bibliographystyle{IEEEtran}

\bibliography{bib}

\end{document}